\documentclass[apj]{emulateapj}
\usepackage{natbib,graphicx,apjfonts}
\usepackage{natbib,graphicx,apjfonts}
\submitted{}
\journalinfo{Accepted for publication in \apj}

\newcommand{\po}{\varpi}

\newcommand{\fij}{\varphi_j}
\newcommand{\dfij}{\dot{\varphi}_j}

\newcommand{\Cs}{C_{s1}}
\newcommand{\Css}{C_{s2}}
\newcommand{\tCr}{\tilde{C_r}}

\def\be{\begin{equation}}
\def\ee{\end{equation}}
\def\ba{\begin{eqnarray}}
\def\ea{\end{eqnarray}}

\shorttitle{Binary System Stability}
\shortauthors{Mudryk and Wu}

\begin{document}

\title{Resonance Overlap is Responsible for Ejecting Planets in Binary Systems}

\author{Lawrence R. Mudryk and Yanqin Wu}
\affil{Department of Astronomy and Astrophysics, University of Toronto,
    Toronto, ON M5S 3H8}
\email{mudryk@astro.utoronto.ca}

\begin{abstract}
A planet orbiting around a star in a binary system can be ejected if
it lies too far from its host star. We find that instability boundaries
first obtained in numerical studies can be explained by overlap between 
sub-resonances within mean-motion resonances (mostly of the j:1 type). 
Strong secular forcing from the companion displaces the centroids of 
different sub-resonances, producing large regions of resonance overlap. 
Planets lying within these overlapping regions experience chaotic 
diffusion, which in most cases leads to their eventual ejection.  The 
overlap region extends to shorter-period orbits as either the companion's 
mass or its eccentricity increase. Our analytical calculations reproduce 
the instability boundaries observed in numerical studies and yield the 
following two additional results. Firstly, the instability boundary as a 
function of eccentricity is jagged; thus, the widest stable orbit could be 
reduced from previously quoted values by as much as $20\%$. Secondly, very 
high order resonances (e.g., 50:3) do not significantly modify the 
instability boundary despite the fact that these weak resonances can produce 
slow chaotic diffusion which may be missed by finite-duration numerical 
integrations. We present some numerical evidence for the first result. 
More extensive experiments are called for to confirm these conclusions. 
For the special case of circular binaries, we find that the Hill criterion 
(based on the critical Jacobi integral) yields an instability boundary 
that is very similar to that obtained by resonance overlap arguments, 
making the former both a necessary and a sufficient condition for planet 
instability.
\end{abstract}


\section{Introduction}

The majority of solar-type stars in our neighborhood ($\sim 60\%$)
are in binary or higher-multiple systems
% the actual ratios of system types is given as single:binary:triple:quardruple = 57:38:4:1
% so 60% of stars are in binary systems or higher
\citep{mayor91}. Despite this majority, there are still questions 
as to how many of these multiple-star systems host planets, and
whether or not the planet formation process inside these systems
differs markedly from that around single stars. Radial-velocity
surveys have shown that $\sim 20\%$ of the extra-solar planets reside
in binaries \citep{eggenberger04}, but the true fraction is likely
higher as these surveys select against observing known binaries.

While it is clear that much theoretical and observational effort is
still needed to fully answer the above questions, much progress has
been made in one sub-area of this issue---the dynamical stability of
planets in binary systems. The body of literature on this topic is
extensive, with most studies using numerical techniques.
\citet{henon} numerically studied periodic planet orbits 
in circular binaries (circular restricted problem)
as a function of the binary mass ratio.
\citet{benest93} included binary eccentricity in his study but only focused on 
a few astronomical systems. \citet{rabl} also considered eccentric
binaries but limited their studies to equal-mass
stars. \citet[hereafter HW99]{hol99} is the most comprehensive and
homogeneous study to date. They numerically integrated (initially
circular) planet orbits for $10^4$ binary periods, and charted out the
stability region as a function of binary separation, eccentricity, and
mass ratio, for both the S-type (circum-stellar) and P-type
(circum-binary) planetary orbits.
\citet{dvorak02} have since included the effect of planet eccentricity but
found it to be less important than the binary eccentricity.
With the intention to quantify the confines of habitable zones around
binary stars, \citet{musielak05} also investigated the stability of both 
S-type and P-type planetary orbits in circular binaries.  To this end, 
they adopted a criterion for stability that differs slightly from that 
used in other works. However, they found results that largely agree with
those from previous works, including those of HW99. 
\citet{marzari05} examined the stability of multiple planets in
binary systems, including the effects of mutual planetary
perturbations.  In this case, interactions among the planets
themselves appear to be the leading cause for instability.
\citet{david03} concentrated on studying ejection timescales for planets 
within the unstable region.  They established an empirical formula for the
ejection timescale that is a steep function of the periastron distance
for the binary companion.  Beyond a certain distance, however, this
trend is expected to break down and the ejection time become infinite (the
system becomes stable). The location of this break is the boundary for
which we are interested in searching. Since the afore-mentioned papers
have confirmed the HW99 results, we focus on comparing our analytical
results against those of HW99 exclusively. 

The numerical results of HW99 and \citet{rabl} uphold the expectation
that the stability space (comprising the binary's eccentricity and the
ratio of the planet's semi-major axis to that of the binary) shrinks
with decreasing stellar separation, with increasing orbital
eccentricity, and with increasing companion mass.  However, the
underlying physical mechanism for planet ejection has yet to be
demonstrated.  Moreover, current computational capabilities limit the
integration time (up to $10^4$ binary orbits in HW99) and permit only
coarse-grid parameter searches. The former limitation may allow
longer-term instabilities to be missed while the latter blurs the
transition from stability to instability, hiding the existence of
possible \mbox{`(in-)stability} islands.'

Our current study aims to overcome these limitations. We expose the
instability mechanism, delineate the topology of transition between
stability and instability,
and exclude the possible existence of longer-term instabilities.  We
accomplish these aims by studying individual orbital resonances and
the conditions for which they overlap, adopting resonance overlap as a
necessary and sufficient criterion for chaotic diffusion, and
consequentially, for planet instability. In this work, we focus our
attention on the orbits most relevant for radial-velocity searches---the
so-called S-type orbits \citep{dvorak84}, where the planet orbits
around one of the stars. The second star is considered to be an
external perturber. We also limit our studies, like most numerical
works, to coplanar systems.
Non-coplanarity introduces new resonances, which may render the systems
more unstable.

In this paper, we first present analytical arguments that allow us to
determine the boundaries of stability (\S \ref{section:formalism})
and then compare them against numerical results from HW99 (\S
\ref{section:overlap_results}). We dedicate a special section to the 
case of circular restricted problem (\S \ref{sec:circular}) and
present our conclusions in \S \ref{section:conclusions}.

\section{Resonance Overlap} \label{section:formalism}

Chaotic diffusion associated with resonance overlap has been shown to
be responsible for all known cases
of instability in the solar system, including the clearing of the
Kirkwood gaps within the asteroid belt and the orbital stability of
planets and short-period comets
\citep{wisdom,der83}. In this context, the most massive perturber,
Jupiter, has a mass ratio to the Sun of $\sim 10^{-3}$, and in
addition to (fairly low-order) 2-body mean-motion resonances, both
secular resonances and 3-body resonances have also been shown to be
relevant \citep[for a review, see][]{lecar01}.

The situation in a binary system is different. The companion star is
the lone perturber with a mass ratio of order unity. The only relevant
resonances are the 2-body mean-motion resonances (MMRs), but these are
not limited only to low-order.

Despite having $\mu = m'/(m_c + m') \sim 1$ in our case, where $m'$ is the
companion mass and $m_c$ the host star mass, we adopt the formalism of
the disturbing function formally derived for $\mu
\ll 1$. We believe that this formalism includes all relevant resonant angles
and correctly describes the resonance strength to order unity, which
is sufficient for our purpose (more later).
The disturbing function for the planetary orbit has the form
\citep[][hereafter MD99]{cdmurray99}
% eq. (6.137)
\begin{equation}
{\cal R} = {{G m'}\over{a'}}\, \sum S_j \cos \fij,
\label{eq:disturbing}
\end{equation}
where $a'$ is the semi-major axis of the binary companion. We use
primed variables for the orbital elements of the binary companion, and
unprimed variables for those of the planet (which has mass $m$). $S_j$ is a
coefficient that depends on the eccentricities of the planet and the
external companion, and on the ratio of their semi-major axes, $\alpha =
a/a'$. The mean motion, $n$, is expressed as $n^2 = G(m_c + m)/a^3$.
The angle argument is
\begin{eqnarray}
\fij = j_1 \lambda' + j_2 \lambda + j_3 \po' + j_4 \po,
\label{eq:phitext}
\end{eqnarray}
where $\po$ is the longitude of pericentre, and $\lambda$ is the mean
longitude. The mean longitude and mean motion are related by $\lambda
= nt + \epsilon$ where $\epsilon$ is the mean longitude at
epoch. Similar relations exist for the companion. The summation in
equation (\ref{eq:disturbing}) is formally over all integer
combinations $(j_1, j_2,j_3,j_4)$ that satisfy the {\it d'Alembert
relation}: $j_1 + j_2 + j_3 + j_4 = 0$. However, at a given $\alpha$
value, only a few combinations are relevant for the dynamics---for the
other combinations, the angle $\fij$ varies with time too fast to have
any sustained long-term effect.  Removing these fast oscillating terms
by integrating over an appropriately long time and keeping terms to
the lowest order in eccentricities, we obtain the averaged disturbing
function,
\begin{eqnarray}
{\cal R} & = & {{G m'}\over{a'}} \left[ f_{s1} (e^2 + e'^2) + f_{s2} e
e' \cos(\varpi' - \varpi) \right. \nonumber \\
& & \left.  + f_r e'^{|j_3|} e^{|j_4|} \cos \fij\right].
\label{eq:disturbing2}
\end{eqnarray}
The first two terms in the brackets arise from the two lowest order
secular interactions,
%($j_1 = j_2 = 0$, $j_3 = j_4 =0$ or $j_3 = - j_4 = 1$), 
while the last term accounts for MMRs situated at $j_1 n' + j_2 n
\approx 0$. In particular, these could include resonances that share
the same $j_1$ and $j_2$ values but have different $j_3$ (and
therefore $j_4$) values. We call these `sub-resonances' of a given MMR
$(j_1,j_2)$.  Their importance will become clear later.  The
coefficients $f_{s1}$ and $f_{s2}$ are functions of $\alpha$
alone. Explicit expressions for them are presented in Table B.3 of
MD99.
The $f_r$ coefficient depends on $\alpha$ as well as the particular
resonance under consideration. MD99 cited two expansion formulas
(eqs.\ [6.36] and [6.113] in MD99) to calculate the interaction
strength for any resonance and listed explicit expressions for
low-order resonances (Appendix B of MD99).
We find that both expansion formulas yield similar results and agree 
with the explicit formula at low-order. These expansions
diverge for $e' > 0.6627434$ (see MD99), so we limit our studies to $e' <
0.6$.

Variations of the planet's orbital elements are obtained using
Lagrange's equations as
\begin{eqnarray}
\dot{n}   & = & 3 j_2 C_rne'^{|j_3|}e^{|j_4|}\sin\fij, \label{eq:ndot} \\
\dot{e}   & = & -\Css e'\sin(\Delta\varpi)  + j_4 C_r e'^{|j_3|}e^{|j_4|-1}\sin\fij, \label{eq:edot} \\
\dot{\po} & = &   2\Cs   + \Css \frac{e'}{e}\cos(\Delta\varpi) + |j_4| C_r e'^{|j_3|}e^{|j_4|-2}\cos\fij, \label{eq:podot}\\ 
\dot{\epsilon}& = &\Cs e^2 +{{\Css}\over 2} e'e\cos(\Delta\varpi)+{{|j_4|}\over{2}}C_r e'^{|j_3|}e^{|j_4|}\cos\fij, \label{eq:epsdot}
\end{eqnarray}
where $\Delta \varpi$ is the secular angle argument, $\Delta \varpi =
\varpi' - \varpi$. Moreover, the $C$-coefficients are related to the
$f$-coefficients in equation (\ref{eq:disturbing2}) by $C_x =
[Gm'/(na^2a')] f_x \approx (m'/m_c) n \alpha f_x (\alpha)$.  The
variation in $\epsilon$ is smaller than that in $\varpi$ by a factor
of $e^2$ and can be neglected. Perturbations of the companion's
orbital elements due to the planet are also ignored.

Exact resonance occurs when both $\sin\fij = 0$ and $\dot {\fij} =
0$, \textit{viz}.\
$j_1 n' + j_2 n + j_4 {\dot \varpi} = 0$. Near this location, $\fij$
librates about the resonant value.
Moving away from this location, there exists a boundary beyond which
$\fij$ changes from libration to circulation. This boundary defines
the \textit{width} of the resonance, namely, the range of space over
which the resonance dominates the planet's dynamics.
When the widths of two resonances become comparable to their
separation, the planet can be affected simultaneously by these
overlapping resonances.
Mathematically, the overlap of two or more resonances causes
neighboring trajectories to diverge exponentially with time 
\citep{chirikov79, wisdom}. This occurs on the Lyapunov timescale ($T_L$), 
which, as argued by \citet{holman96}, is comparable to the libration
timescale for the resonances in question.  Except in the case of
`bounded chaos,' orbital parameters for the planet undergo unbounded
random walks leading to ejection on a timescale called the event
timescale ($T_e$). Though $T_e$ fluctuates depending on the system,
studies have shown that it roughly correlates with $T_L$ \citep[see,
e.g.][]{lecar92}.

In our problem, the large companion mass produces strong secular
forcing, making it different from typical solar-system dynamics
problems.  Firstly, even if the planet initially has zero
eccentricity, it is forced to oscillate with an eccentricity amplitude
(eqs.\ [\ref{eq:edot}] and [\ref{eq:podot}])
\begin{equation}
e_{\rm sec} = {{\Css} \over{2 \Cs}}\, e'\,,
\label{eq:esec}
\end{equation}
on the short secular timescale $\Delta t \sim
2\pi/(2\Cs)$.  The magnitude of $e_{\rm sec}$ decreases with
decreasing $\alpha$ but can be as large as $e'/2$ near the 3:1
resonance. We find that $e_{\rm sec} \sim \alpha e'$. For equal-mass
binaries, the secular timescale ranges from $\sim 20$ planet orbital
periods at the 3:1 resonance to $\sim 1000 $ periods at the 20:1
resonance.
While the secular timescale is likely too long when compared with the
mean-motion of the companion to allow for the `evection resonance'
\citep{touma},\footnote{When the companion mass dominates, $\mu \rightarrow 1$, 
the evection resonance may become important. See \S
\ref{sec:circular}.} it is typically much shorter than the resonant
timescale.  Considering also that the resonant strength is at maximum
when the value of $e$ is at its largest, we can assume that $e = e_{\rm sec}$ for the
resonant interactions.  Planets possessing a free eccentricity in
addition to the forced value can reach higher overall eccentricity
and will therefore be more unstable.

The second effect of the companion's strong secular forcing is to
displace the centroid of different sub-resonances away from each
other. In Appendix \ref{section:A1}, we derive expressions for the
centroid and the width of a resonance when secular forcing is
important. While the width remains unchanged from the non-secular
case, the centroid of a MMR is shifted from its nominal position, ($j_1
n' + j_2 n =0$), by an amount, $|\delta n|
\sim |2 (j_4/j_2) \Cs|$. As a result, the region of resonance overlap
is greatly expanded.  This effect is illustrated in Fig.\ \ref{fig:subresonance}
for two groups of MMRs. 

Resonance overlap generates chaotic diffusion, but as pointed out
by \citet{murray92}, under some circumstances resonance overlap will 
lead only to `bounded chaos'---unpredictable but limited variations
in the orbital elements. One such example is provided by \citet{gladman}. 
Results from our numerical experiments (Fig.\ \ref{fig:density_9}) 
as well as discussions in \S \ref{sec:circular} suggest that this is not a major
concern for determining the instability boundary. In the remaining discussion we, therefore, 
do not distinguish between the concepts of resonance overlap and planet
instability.

Another question relates to whether the overlap between sub-resonances
is as potent as that between distinct MMRs, thereby leading to planet
instability on an astronomically interesting timescale \citep[see the
review by][]{malhotra}.
We present calculations in \S \ref{section:overlap_results} which 
suggest that this is indeed so.
%%%%
%the chaotic diffusion caused by the sub-resonance overlap leads to planet
%ejection within 1 Gyr (often much faster) for most typical systems.  
%%%%

\section{Comparison with HW99 and Discussion}
\label{section:overlap_results}

In our determination of regions of resonance overlap, we include
resonances with $j_1 \geq 3$, $ -4 \leq j_2 \leq -1$, and $- |j_1 +
j_2| \leq j_3 \leq 0$. We restrict the value of $j_2$ since the
strength of a resonance scales as $e'^{|j_3|}\, e^{|j_4|} \propto
e'^{|j_1 + j_2|}$ (eq.\ [\ref{eq:disturbing2}]). For a given orbital
separation, $\alpha$, the ratio of $j_2/j_1$ is determined by Kepler's
third law: $\alpha^3 = (j_2/j_1)^2 (1-\mu)$; hence, the strongest
resonances have $j_2 = -1$. In fact, we show that even the $j_2 = -4,
-3$, and $-2$ resonances do not affect the instability boundary much.
Moreover, while $j_3 = 0$ is the strongest sub-resonance in solar
system dynamics (in light of Jupiter's small eccentricity), we find
here that all sub-resonances are essential to determine the overlap
region.

\begin{figure}
\epsscale{1.2}
\plotone{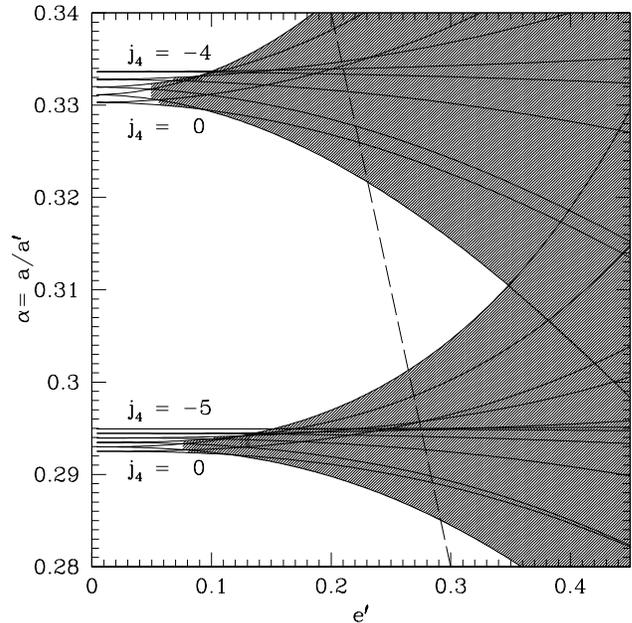}
\caption{
Location and width of various sub-resonances as a function of $e'$,
obtained for a $\mu = 0.1$ binary. The top group are the 
sub-resonances of the 5:1 MMR (identified by their respective $j_4$
values) and the lower group, of the 6:1 MMR. We take the planet
eccentricity to be the secularly-forced value.  The centroids of
different sub-resonances within a distinct MMR are displaced from each
other due to both secular and resonant forcing though the secular
effect dominates at low values of $e'$. Shaded regions are regions of
instability, as defined in the text.  Overlap between sub-resonances
of the same MMR covers a much larger region than overlap between
distinct MMRs.}
\label{fig:subresonance}
\end{figure}

Coupling strengths are calculated using the afore-mentioned series
expansion formulas in MD99 (eqs.\ [6.36] or [6.113]). The location
and width of each resonance are obtained as described in Appendix
\ref{section:A1}. Planet eccentricity is taken to be the secularly-forced 
value (eq.\ [\ref{eq:esec}]).  All coefficients are evaluated at
exact resonance, assuming the resonance width is small. In the
$(\alpha,e')$ phase space, a region is designated as unstable if more
than one resonance (or sub-resonance) spans it. We further assign
a similar status, at the same value of $e'$, to the entire extent in $\alpha$ of the 
sub-resonance in which this region is situated (see Fig.\ \ref{fig:subresonance}). 
Depending on its orbital phase, a planet situated within a single resonance (elsewhere spanned by additional resonances),
but which is still outside of the region of overlap proper, may (or may not) librate 
into the latter. This definition of unstable regions ensures that all 
potentially unstable orbits are included.  Again, our analytical study 
is limited to $e' < 0.6$ to ensure a converging disturbing function.

\begin{figure}
\begin{center}
\epsscale{1.2}
\plotone{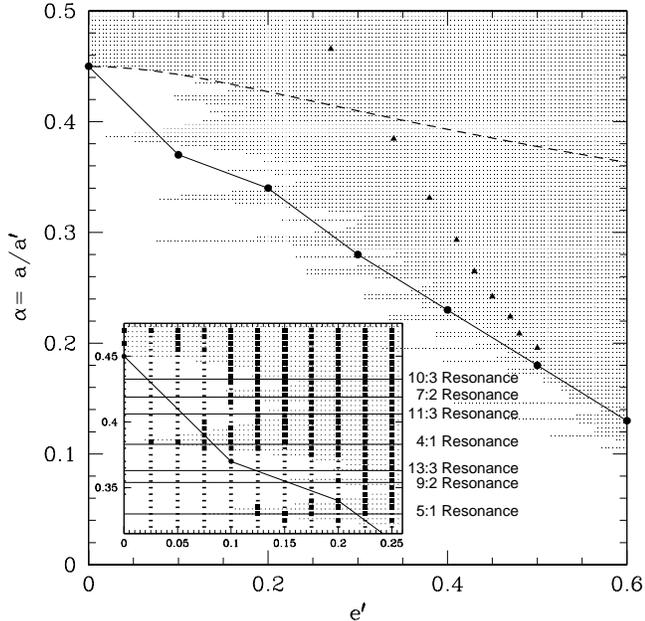}
\caption{
Stability diagram for planets in a $\mu=0.1$ binary system. 
The solid curve connecting filled circles locates the
maximum stable value of $\alpha = a/a'$ as obtained by HW99 while dots map
regions of instability caused by resonance overlap.
%we find two or more resonances overlap. Resonances
Resonances included in this calculation are described in the text.
The instability boundary as it exists when considering only the distinct MMRs
(keeping $j_3 = 0$) is denoted by filled triangles. Over the
eccentricity range of interest, overlap between sub-resonances is the
most significant source of planet instability.  As $e' \rightarrow 0$,
widths of most resonances approach 0 except for the 2:1 and 3:1
resonances. The dashed curve shows the lower confine of a 3:1
resonance---overlap between sub-resonances within the 3:1 MMR can
explain instability in circular binaries.
{\bf Inset:} results of numerical integration over a selected region
of $\alpha-e'$ space. Dashes represent planet orbits that remain stable for more
than $3000$ binary periods; filled squares represent unstable
orbits. Horizontal lines indicate centroid locations of certain MMRs
that are responsible for the jutting peninsulas. At each $e'$ value,
$j_2=-1$ MMRs yield the shortest-period unstable orbits.
Stability for points near the instability boundary are sensitive to
the initial conditions.
Regions of resonance overlap coincide with that for planet
instability and there is little evidence for `bounded chaos.'
\label{fig:density_9}}
\end{center}
\end{figure}

\begin{figure}
\begin{center}
\epsscale{1.2}
\plotone{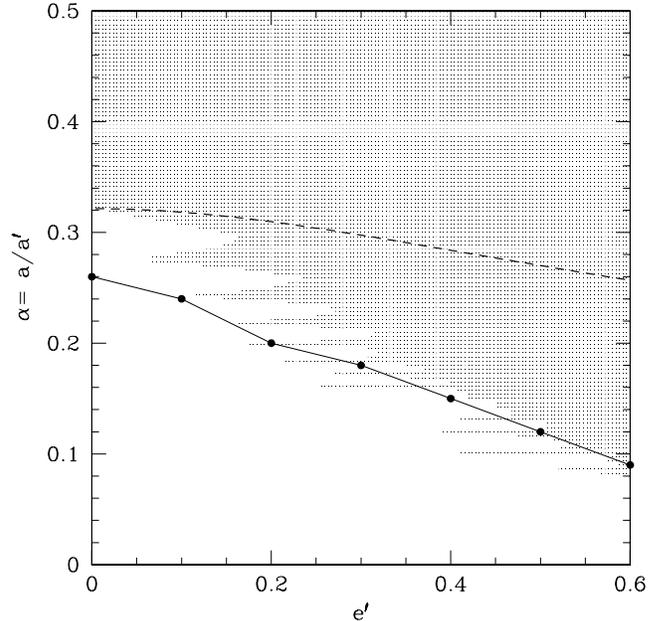}
\caption{
Stability diagram for planets in an equal-mass binary system ($\mu =
0.5$). Symbols are the same as those in Fig.\ \ref{fig:density_9}. 
We obtain these results using a perturbation formula that is strictly
valid only for $\mu \ll 1$---this may account for some of the
discrepancy between our results (dots) and those of HW99 (solid curve).
\label{fig:density_eq}}
\end{center}
\end{figure}

Our full results are shown in Fig.\
\ref{fig:density_9} for mass-ratio $\mu = 0.1$,
and in Fig.\ \ref{fig:density_eq} for $\mu = 0.5$. As has been
indicated in Fig.\ \ref{fig:subresonance}, the instability boundary is
jagged, with jutting peninsulas and narrow inlets. This is different
from the smooth lines presented by HW99. However, their curves largely
trace the outline of our results. The two sets of results can be
considered consistent since HW99 carried out their investigation over 
a crude grid in $\alpha-e'$ space. To confirm this,
we perform similar numerical integrations, with a much finer grid in a
selected region of $\alpha-e'$ space. We adopt the Hierarchical Jacobi
Symplectic integrator by \citet{beust}, an add-on to the SWIFT package
\citep{duncan98} for studying dynamics in multiple-star stellar
systems. Planets are initialized to have random orbital phases and an
eccentricity given by $e=e_{\rm sec}$ (initializing planets with zero
eccentricity produces similar results). We integrate their orbits for
$3000$ binary periods. The stability of these orbits is indicated in
the inset of Fig.\ \ref{fig:density_9}. The detailed topology agrees
well with that obtained from our perturbation analysis, and in many
cases, one can even identify the (sub)-resonances responsible for the
instability.
This suggests that resonance overlap and the consequential chaotic
diffusion is the mechanism responsible for the planet instability
observed in HW99's numerical investigation. Moreover, there is little
evidence for `bounded chaos' near the instability boundary, so that one
can adopt the boundary of resonance overlap as the boundary for planet instability.

\begin{figure}
\begin{center}
\epsscale{1.2}
\plotone{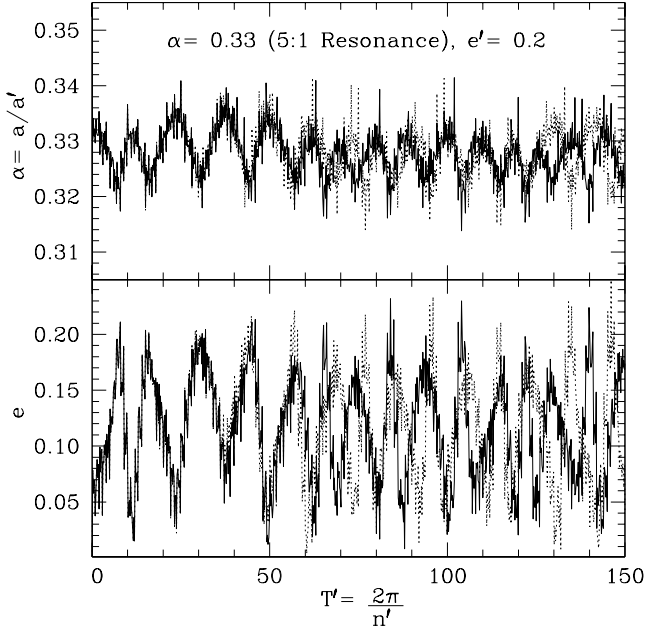}
\caption{
Numerical integration of two planets initialized with identical
orbital parameters except for a tenth of a percent difference in the
initial orbital phase. The resulting semi-major axis and eccentricity
are plotted as functions of time (measured in binary orbital periods,
$T'$) in solid (or dotted) curves for each planet. The Lyapunov
timescale is estimated to be $\sim 10$ binary orbits, and
de-correlation in the semi-major axis and eccentricity becomes
apparent to the eye after approximately $50$ binary orbits.  The planets
are ejected in turn after about $1500$ and $3800$ orbits.
\label{fig:chaotic_diffusion}}
\end{center}
\end{figure}

In an effort to delineate the differences between the chaotic dynamics existing
within regions of resonance overlap and the regular dynamics existing just outside
such regions, we numerically integrate two sets of two initially `close' planets. 
Both sets of planets are situated near the 5:1 MMR in a binary system with 
mass ratio, $\mu= 0.1$, and eccentricity, $e'=0.2$.  The first set of two 
planets are situated directly within the region of overlap at $\alpha=0.33$, while
the second set are situated just outside the region of overlap at $\alpha=0.32$.
Fig.\ \ref{fig:chaotic_diffusion} presents the results of integrating the first 
set of two planets initialized with identical orbital parameters except for a 
tenth of a degree difference in their orbital phase positions. 
The Lyapunov timescale is defined as the timescale for exponential
divergence between two infinitesimally close orbits. We roughly estimate this
timescale for the trajectories presented in Fig.\ \ref{fig:chaotic_diffusion}
and obtain $T_L \approx 10$ binary orbits.
De-correlation in the semi-major axis and eccentricity becomes
apparent to the eye after approximately $50$ binary orbits. The libration
time within this resonance, which one expects to be of the same order
as the Lyapunov time, is $\sim 54$ binary orbits. The planets are
ejected in turn after about $1500$ and $3800$ binary orbits.  Further
integrations at the same location with different initial orbital
phases show a wide spread in the ejection times ranging from
$50$--$4000$ binary orbits, corresponding to $\sim 10^4$ -- $10^6$
years for a solar-mass binary at $50$ AU.
\citet{murray97} and \citet{david03} presented two different empirical
expressions that relate the (widely scattered) ejection time ($T_e$)
to the Lyapunov timescale. The former found a relationship between
these two timescales (applicable to overlapping sub-resonances) given
by $T_e/T' = 10^a(T_L/T')^b$ where $a=1.45$ and $b=1.68$, and where
$T'$ denotes the binary orbital period. Applying this formula to our
case yields an ejection time of $T_e/T' \sim 2000$. The expression by
\citet{david03} yields a comparable value of $T_e/T' \sim 7400$. Within the scatter,
both values agree with our experiments.

Adopting the expression by \citet{murray97}, we obtain ejection times
for various resonances. Lower order resonances lead to faster
ejection, while at the higher end, for example, the 30:1 resonance, we
find $T_e/T' \leq 10^6$ for a system with mass-ratio, $\mu = 0.1$.
This resonance (corresponding to $\alpha = 0.10$) defines the maximum
stable orbit for the most eccentric binary orbit we consider ($e'=0.66$).
The ejection time corresponds to $\sim 400$ Myrs for a $50$ AU binary
and $\sim 1$ Gyrs for a $100$ AU binary (all assuming a total system
mass equal to one solar-mass). We conclude that overlap of
sub-resonances leads to planet ejection on astronomically interesting
timescales, for the parameters we have considered.

By contrast, the results of the second set of integrations where we
instead situate the two planets just outside the region of resonance
overlap at a location given by $\alpha = 0.32$, do not
exhibit sensitivity on the initial conditions and no planet is ejected
within our integration time ($10^4 T'$). The transition to instability
occurs over a narrow region.
%---a difference of two-hundredths of a degree in the
%orbital phase results in less than half a percent difference between
%the orbital parameters after $10^4$ binary orbits.  

One major discrepancy between our results and those of HW99 can be
observed in Fig.\ \ref{fig:density_eq} for equal-mass binaries: at low
binary eccentricity, the HW99 curve falls below that obtained from our
perturbation analysis. This likely reflects failure of the expansion
formula when $\mu$ is large (more below).

A key question of interest asks, what is the longest-period
stable planet orbit, for a given binary (fixed $\mu$ and $e'$)?  HW99
provided a fitting formula for the minimum unstable $\alpha$ as a function
of $\mu$ and $e'$. Our results here indicate, however, that the
minimum unstable $\alpha$ should be reduced by up to $\sim 20\%$ from
their values. This is related to the thin instability peninsulas
evident in our Figs. \ref{fig:density_9} and
\ref{fig:density_eq}.

\begin{figure}
\epsscale{1.2}
\plotone{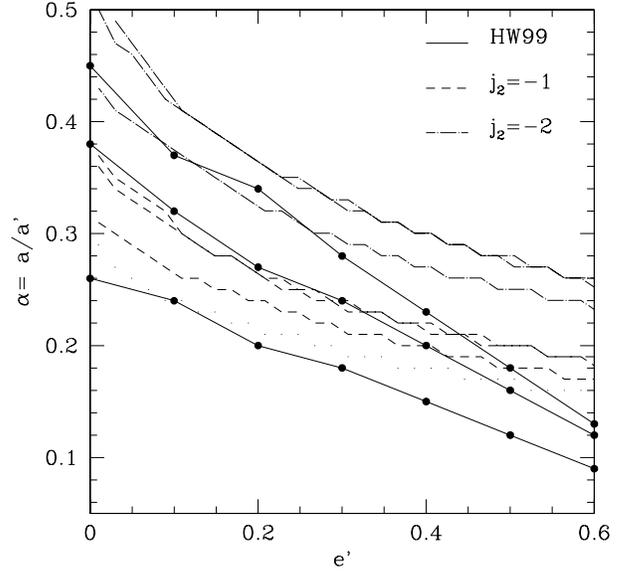}
\caption{
Comparison of instability boundaries obtained based on simple
approximations of our analytical arguments (eq.\ [\ref{eq:crit}]) and
numerical results of HW99.  The group of dashed curves represent
the approximate overlap condition for $j_2 = -1$ MMRs, while the
dot-dashed ones represent those for $j_2 = -2$ MMRs, both with $|j_4| =
j_1/2$. Within each group, from top to bottom, the value of the
mass-ratio is $\mu = 0.1, 0.2$ (these two curves almost coincide), and
$0.5$, respectively.  If we recalculate the bottom-most dashed curve
($\mu = 0.5$) assuming the value of $\Cs/C_r$ is $10$ times smaller,
we obtain the results shown in the dotted curve.
\label{fig:fitting}}
\end{figure}

In order to understand how the outline of the instability boundary
depends on various parameters, we propose the following rough scaling
argument.  Let neighboring sub-resonances be spaced by $\Delta n$,
where due to secular forcing, $\Delta n \approx 2|\Cs/j_2|$ (eq.\
[\ref{eq:finaln0}]). Resonant interactions also modify the
centroid of a resonance, but they are less important than the secular
effect for small $e'$. The width of an individual sub-resonance, $k$,
is expressed in equation (\ref{eq:finalk}), which in most cases can be
simplified as $k \approx [4j_4^2/(3j_2)] \, |C_r| e'^{|j_3|}
e^{|j_4|-2}$.  Adopting $e = e_{\rm sec} \approx
\alpha e'$, and requiring resonance overlap ($\Delta n < 2 k$), we
find that instability occurs when
\begin{equation}
\alpha \geq \alpha_{\rm crit} = 
\left({3\over {4 |j_4|^2 e'^{|j_1 + j_2|-2}}} 
\left|{\Cs\over{C_r}}\right|\right)^{1/(|j_4|-2)}.
\label{eq:crit1}
\end{equation}
Defining $f_4 = |j_4/j_1|$, and relating $j_1$ to $\alpha$ by Kepler's
Law, $|j_1/j_2|^2 \alpha^3 = 1-\mu$, we recast equation
(\ref{eq:crit1}) as
\begin{equation}
\alpha \geq \alpha_{\rm crit} \approx 
\left({3\over {4 f_4^2 |j_2|^2 (1-\mu) e'^{|j_1 + j_2|-2}}} 
\left|{\Cs\over{C_r}}\right|\right)^{1/(f_4|j_1|-5)},
\label{eq:crit}
\end{equation}
where $j_1$ is also a function of $\alpha$.  Numerically, we observe
that, regardless of the mass ratio and the resonance involved,
$|\Cs/C_r|$ rises monotonically with $f_4$ and
clusters around $0.02$ when $f_4 \approx 0.5$. For simplicity, we
solve for $\alpha_{\rm crit}$ considering only $f_4 \approx 0.5$. The
results are plotted in Fig.\ \ref{fig:fitting} for three mass
ratios. When only $j_2 = -1$ MMRs are considered, the $\mu=0.1$ and
$\mu =0.2$ results sit atop each other falling somewhat below the
respective HW99 curves at small values of $e'$ and above them at large
values. Besides errors resulting from our crude approximation in taking
$f_4 = 0.5$, two other factors may contribute to this discrepancy. The first
is that we are searching for the very minimum value of $\alpha$ at each
value of $e'$ that allows resonance overlap. As is shown in
Fig.\ \ref{fig:density_9}, this may lie up to $20\%$ below the HW99
numerical result. The second factor is that we ignore overlap between
distinct MMRs, which may be important at sufficiently large $e'$.
When $\mu=0.5$, our curve consistently sits above the HW99 line,
resembling the discrepancy shown in
Fig.\ \ref{fig:density_eq}. This, as we argue above, likely reflects
failure of the expansion formula when $\mu$ is large (more below).

Despite these short-comings, this simple analysis yields some useful
insight. Comparing overlap conditions between those resonances with
$j_2 = -1$ and those with $j_2 = -2$, reveals that the latter resonances
always occur at a larger value of $\alpha$ for a given $e'$ value.  They are
therefore not important for determining the instability boundary and
we conclude that our neglect of $|j_2| > 4$ MMRs is valid. This point
is further emphasized in the inset of Fig.\ \ref{fig:density_9}. Based
on this conclusion, we argue that instability boundaries obtained from
finite-duration numerical integration are reliable, even though they
may not detect long-term instabilities brought about by very high-order
resonances (e.g., 50:3). A second point concerns the fact that we
have ignored terms of order $\mu^2$ in the expansion of the disturbing
function, and that our $\Cs$ and $C_r$ coefficients are only correct
to order-of-magnitude. We argue, however, that the instability boundary
depends only on the ratio of $\Cs/C_r$ (eq.\ [\ref{eq:crit}]).
Moreover, if, for instance, the true ratio of $\Cs/C_r$ is smaller by
a factor of $10$ than our adopted value of $0.02$, the instability
boundary for $\mu = 0.5$ is moved downward in $\alpha$ by as much as
$10\%$ (Fig.\ \ref{fig:fitting}).

Our arguments here are based on very crude scaling relationships. They
ought to be checked using more elaborate numerical experiments.

\section{Circular, Restricted Three-Body Problem}
\label{sec:circular}

We focus on $e'=0$ binaries to study the following two issues: the
applicability of the Hill criterion for predicting planet instability,
and the relevance of `bounded chaos' that prevents us from using 
resonance overlap as a synonym for planet instability.

\begin{figure}
\epsscale{1.2}
\plotone{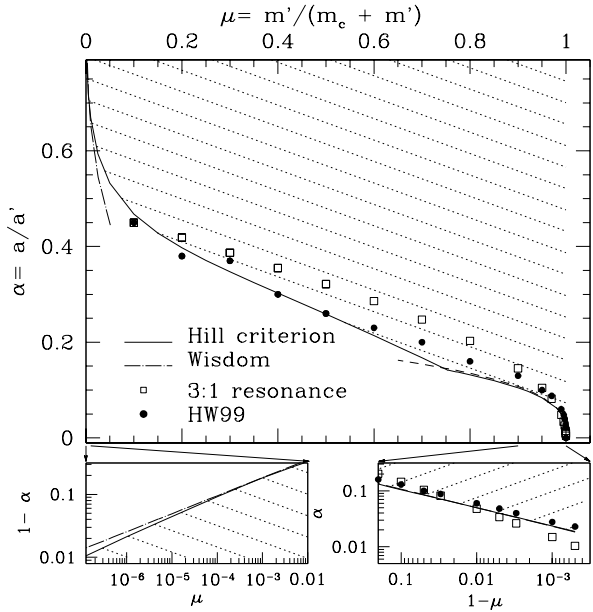}
\caption{
Instability boundary for circular binaries as a function of mass ratio
$\mu$.  The solid curve depicts the result obtained based on the Hill
criterion. Planets situated within shaded regions could potentially be ejected from
the host star, though they will not be unless their orbits are chaotic.
Overplotted are analytical results for the boundary of resonance
overlap (and therefore chaos, but not necessarily ejection): as $\mu
\rightarrow 0$, the overlap condition between $p+1$:$p$ resonances yields
$\alpha_{\rm crit} \geq 1- 1.307 \mu^{2/7}$ \citep{wisdom}; for larger
$\mu$ values, overlap between the $(3,-1,-1,-1)$ and $(3,-1,0,-2)$
sub-resonances occurs for $\alpha$ values above the open squares (this
study). Locations of the open squares are calculated using expansion
formula strictly valid only for $\mu \ll 1$. Also plotted (in filled
circles) are the numerical results by HW99---planets situated above
the filled circles are numerically shown to have unstable orbits.  Left 
and right lower panels expand the view near $\mu= 0$ and $\mu = 1$, respectively.
For $\mu \rightarrow 1$, the Hill criterion is well
quantified as $(1/3) R_H = 0.23 (1-\mu)^{1/3}$ (dashed curve).
\label{fig:restricted2}}
\end{figure}

With the exception of those of first and second-order, the widths
of all other MMRs approach zero in circular binaries (see Appendix
\ref{section:A1}). The resonance overlap condition in this limit 
is particularly easy to analyze.  In the case where $\mu \rightarrow 0$
(the sun-asteroid-planet problem),
\citet{wisdom} derived the overlap condition
between first-order ($p+1$:$p$) resonances as described by $|1-\alpha| =
|a'-a|/a' \leq 1.307 \mu^{2/7}$ \citep[also
see][]{duncan,malhotra}. For larger $\mu$ values, we argue that
overlap between the $(3,-1,-1,-1)$ and $(3,-1,0,-2)$ sub-resonances
defines the lowest $\alpha$ value for which chaos can set in. Note
that we calculate the resonance location and width using expansion
formulas that are strictly valid only for $\mu \ll 1$. We suspect this
approximation may lead to some uncertainty in the results in Fig.\
\ref{fig:restricted2}. Moreover, we have not considered the `evection
resonance', which becomes important as $\mu \rightarrow 1$ \citep{touma,nesvorny}.

The stability of planets in a circular binary can also be studied
using the concept of Hill stability \citep[e.g. see][]{cdmurray99}. In
such systems, there exists an integral of motion, the Jacobi constant,
which defines permitted regions of planetary movement. For a planet
that begins with a circular orbit around one star (as in HW99), there
is a critical value of $\alpha$ below which the zero-velocity curve with the
same Jacobi constant is `closed' and the planet cannot escape. For values of 
$\alpha$ greater than this critical value, the planet is allowed to
escape by the Hill criterion but will {\it not} unless its orbit is
chaotic due to overlapping resonances. In other words, the Hill
criterion is a necessary but not a sufficient condition for planet {\it
instability}.

To calculate $\alpha_{\rm crit}$, one needs to carefully consider the
phrase `begins with a circular orbit.' For $\mu \ll 1$ systems
(analogous to the sun-asteroid-planet problem), it is more appropriate
to actualize this condition by taking the sidereal velocity (velocity
in the binary center-of-mass frame) to be $G m_c/a$;
%$v^2 = Gm_c/a$, 
while for $\mu \rightarrow 1$ systems (analogous to the
planet-satellite-sun problem), the more reasonable approach is to set
the synodic velocity (in the host star's rotating frame) to be $G
m_c/a$.
%$v^2 = G m_c/a$. 
For intermediate $\mu$ values, we adopt the approach that
yields the higher value of $\alpha_{\rm crit}$. The resultant values of
$\alpha_{\rm crit}$ are plotted in Fig.\
\ref{fig:restricted2} as a function of $\mu$. In particular, in the limit where $\mu
\rightarrow 1$, we find that $\alpha_{\rm crit} = (1/3) R_H' = [(1-\mu)/81]^{1/3}$ 
where $R_H'$ is the Hill radius of the binary companion \citep{szebehely}, and in 
the limit where $\mu \ll 1$, we find that $\alpha_{\rm crit} = 1 - 2.1 \mu^{1/3}$. 

While the Hill criterion gives the energetic condition for planet
instability, resonance overlap provides the dynamical cause. How do
results from the Hill criterion compare with those from the resonance
overlap criterion?  Intriguingly, they seem to closely trace each
other over small $\mu$, intermediate $\mu$ and large $\mu$ values
(Fig.\ \ref{fig:restricted2}). The only exception is when $\mu
\rightarrow 0$ (visible when $\mu \leq 10^{-4}$) for which the
resonance overlap occurs over a larger range than does the Hill
criterion. \citet{gladman} has studied this limit and concluded that
`bounded chaos,' producing unpredictable but limited variations in
the orbital elements, reigns in the intervening region.  This instance,
however, is the only clear sign of bounded chaos in the circular,
restricted problem. 

Numerical results by HW99 (filled circles in Fig.\
\ref{fig:restricted2}) and our own simulations also
confirm this seemingly coincidental agreement between the resonance
overlap condition and the Hill criterion.
It appears then, that in practice, the Hill criterion is not only a
necessary, but also a sufficient condition for planet instability.

%Still, why is this so? Why should two sets of independent arguments
%give rise to the same instability boundary? More investigation is
%warranted.

\section{Conclusions} \label{section:conclusions}

A planet in a binary system experiences both secular and resonant
perturbations from the binary companion. It may be dislodged from its
host star if it is simultaneously affected by two resonances. We find
that overlap between sub-resonances of the same MMR accounts for the
instability observed by HW99 and our own numerical integration. 
There is little evidence for `bounded chaos' and the word `resonance
overlap' can be interchanged with the word `orbital instability'.  Our
instability boundaries largely agree with those obtained by HW99,
albeit with many fine features. The jutting peninsulas and deep inlets
in the instability boundary correspond to the instability (or
stability) islands first observed by HW99. The presence of these
islands suggests that the longest-period stable orbit at each $e'$
value could be reduced by as much as $20\%$ from the HW99 value.
Moreover, our analysis suggests that overlap between very high-order
resonances (e.g., 50:3) do not substantially modify the instability
boundary: these weak resonances, while producing slow chaotic
diffusion, which may be missed by finite-duration numerical
integrations, do not contribute markedly to planet instability.

In detail, the centroids of different sub-resonances are displaced
from each other by the strong secular forcing of the companion
enlarging the phase space of resonance overlap. Chaotic diffusion
caused by sub-resonance overlap is observed to be fast, unlike cases
in the solar system. 
The longest ejection timescale in our study, corresponding to
sub-resonance overlap within the 30:1 MMR, is $\sim 10^6$ binary orbits,
or, $1$ Gyrs for a $100$ AU solar-mass binary. For comparison, the
5:1 MMR overlap gives rise to an ejection time $\sim 2000$ binary
orbits.
% This , which corresponds to the smallest unstable
%$\alpha$ (occurring for $e'=0.6$ binary),%%%%%%%
%leading to planet ejection within 1 Gyr (and often within 10 Myr) for 
%typical systems.

Compared with numerical integrations, our perturbation analysis has the
following short-comings: the perturbation strength is calculated 
accurate only to first-order in the mass-ratio between the companion and
the host star, and the perturbation formula diverges for $e' > 0.66$.

As a final note, we raise the issue of stability in circular binary
systems ($e'=0$). While the Hill criterion (critical Jacobi constant)
gives the energetic condition for planet instability, resonance
overlap provides the dynamical cause. We observe that over almost the
entire range of mass-ratio, the Hill criterion and resonance overlap
yield similar critical $\alpha$ values, making the Hill criterion not
only a necessary, but also a sufficient condition for planet
instability.

\acknowledgments

We thank M. Holman and S. Tremaine for helpful discussions, and an
anonymous referee whose comments helped to improve the paper.  This
research was supported in part by the Natural Sciences and Engineering
Research Council of Canada, as well as the National Science Foundation
of the US under Grant No. PHY99-07949.

\appendix

\section{Width of a mean-motion resonance under secular forcing} 
\label{section:A1}

MD99 have presented a derivation for the width of a MMR when the
resonance angle evolves due to a single resonance. In
our situation with a massive third body, secular effects on the
resonance angle have to be taken into account. We show here how this
modifies the resonance width and resonant centroid.

The relevant resonance angle as well as its time derivatives are,
\begin{eqnarray}
\fij &=& j_1\lambda' +j_2\lambda +j_3\varpi' +j_4\varpi , \label{eq:phi} \\
\dfij &=& j_1 n' +j_2 n +j_4 \dot{\varpi}, \label{eq:phidot}\\
\ddot{\fij} &=& j_2 \dot{n} +j_4\ddot{\varpi}.\label{eq:phiddot}
\end{eqnarray}

The time-variations of $n'$, $e'$, $\po'$, and $\epsilon'$ due to the
influence of the planet are neglected as the planet can effectively be
thought of as a test mass ($m/m_c \ll 1$). We also neglect variations
in $\epsilon$ as previously mentioned.

We take the time-derivative of equation (\ref{eq:podot}), substitute
equations (\ref{eq:edot}) and (\ref{eq:podot}) into the right-hand side,
and use the resulting equations to recast equation
(\ref{eq:phiddot}) into the form
\begin{equation}
\ddot{\fij}  =  \left [3j_2^2\tCr ne^{|j_4|} + |j_4|^2\tCr e^{|j_4|-2} (j_1n'+j_2n) 
- 2|j_4|^3\Cs\tCr e^{|j_4|-2} \right] \sin\fij - |j_4|^3\tCr^2 e^{2|j_4|-4}\sin 2\fij \label{eq:phiddot2},
\end{equation}
where $\tCr = C_r e'^{|j_3|}$. This reduces to equation (8.63) of
MD99 when $|j_4| = 1$ and $\Cs = 0$.
In deriving this equation, we have made some simplifying
assumptions. Firstly, we have ignored the time-dependence of $\Cs$
and $C_r$, which are in reality both functions of $\alpha$. Secondly, we
have neglected the $\Css$ terms in equations (\ref{eq:edot}) and (\ref{eq:podot})
as we expect their time-averaged contributions to be negligible. 

We look for a solution of the system that is pendulum-like, as in the
case without secular forcing. Following MD99, we write $n = n_0 +
k\cos [\fij$/2], where $n_0$ is the mean motion associated with
the nominal value of the resonance and $k$ is a constant that
describes the amplitude of the oscillation, or equivalently, the width
of the resonance. The choice of the angular form, $\cos [\fij/2]$, is determined by the libration amplitude of the resonant angle
$\fij$ ($- \pi$ to $\pi$) as well as the presumed angle where maximum
change in the mean motion occurs ($\fij = 0$). The latter applies when
$C_r < 0 $ and shifts to $\fij = \pi$ when $C_r > 0$; however, the
final result does not depend on the presumed sign of $C_r$.

The derivation that follows is analogous to that presented in
\S 8.7 of MD99. We do not repeat the details here but simply
outline the results:
\begin{eqnarray}
j_1 n' + j_2 n_0 & = & 2 |j_4| \Cs + j_4^2 |\tCr| e^{|j_4| - 2},
\label{eq:finaln0}\\
k & = &  {-2 \over {3}}{ j_4^2\over j_2}| \tCr| e^{|j_4| - 2} \pm 
\sqrt{12 |\tCr| n e^{|j_4|}} \, \left( 1 + {{j_4^4 |\tCr| e^{|j_4| - 4}}
\over{27 j_2^2 n}} \right)^{1/2}.
\label{eq:finalk}
\end{eqnarray}
The secular term is important for shifting the centroid of the
resonance, but does not contribute to the width of the resonance.  In
fact, the width formula is identical to equation (8.75) of MD99 where
secular forcing is ignored.

The simple pendulum approach applies only when the resonant width is
small, i.e., $\delta n = n_{\rm max} - n_{\rm min}
\ll n$. 
Moreover, assuming that $e$ is driven by the secular interaction to a
value that is proportional to $e'$ (eq.\ [\ref{eq:esec}]), most MMRs
have widths which approach 0 as $e' \rightarrow 0$. The first-order
(e.g., 2:1) and second-order (e.g., 3:1) resonances that satisfy $j_4
\neq 0$ are exceptions;
the width diverges for the former and approaches a constant for the
latter.

\end{document}